\def\BEq{\begin{equation}}
\def\EEq{\end{equation}}
\def\BEqA{\begin{eqnarray}}
\def\EEqA{\end{eqnarray}}

\font\bb=msbm10 at 10pt
\font\bbh=msbm10 at 24pt

\font\cal=cmsy10 at 9pt
\font\cals=cmsy10 at 8pt
\font\csc=cmcsc10 at 10pt

\font\rms=cmr10 at 7pt

\font\sf=cmss10 at 10pt 

\def\0#1{\mbox{\rm#1}}
\def\1#1{\mbox{\bb#1}}
\def\2#1{{\mathbf#1}}
\def\3#1{{\cal #1}}
\def\4#1{\mbox{\cals#1}}
\def\5#1{\mbox{\sf#1}} 
\def\6#1{\mbox{\rms #1}}
\def\7#1{\mbox{\bbh #1}}
\def\8#1{{\tilde #1}}
\def\9#1{{\breve #1}}

\def\BE{\begin{equation}}
\def\EE{\end{equation}}
\def\BEA{\begin{eqnarray}}
\def\EEA{\end{eqnarray}}
\def\BEn{\begin{enumerate}}
\def\EEn{\end{enumerate}}

\def\Chi{\mathrm{X}}

\def\Ep{\mathrm{E}}

\def\ga{\gamma}

\def\Si{\Sigma}

\def\om{\omega}

\def\dag{\dagger}

\def\ox{\otimes}

\def\pa{\partial}

\def\from{\kern-2pt\leftarrow\kern-2pt}

\def\x{\times}

\def\Bar{{\Big |}}
\def\bra{\langle}

\def\ket{\rangle}

\def\lo{\stackrel{<}{{}_\sim}}

\def\Cliff{\mathop{\hbox{\rm Cliff}}\nolimits}
\def\CLIFF{\mathop{\hbox{\csc Cliff}}\nolimits}

\def\Dim{\mathop{{\mathrm {Dim}}}\nolimits}

\def\SO{\mathop{\mbox{\rm SO}}\nolimits}

\def\SU{{\mbox{\rm SU}}}

\def\wh{\widehat}

\def\Bar{\kern5pt{\rule[-2.5pt]{.6pt}{9.5pt}}\kern5pt}

\def\GeV{\mbox{ \rm GeV}}

\def\lmult{{\lfloor\kern-5pt\lfloor}}
\def\rmult{{\rfloor\kern-5pt\rfloor}}

\documentclass[]{article}[12pt]
\usepackage{latexsym}
\usepackage{doublespace}

\begin{document}


\title{ ULTRAQUANTUM 
DYNAMICS%
\footnote
{Based on  a talk presented
to IARD 2002, The Third Biennial Meeting of the
International 
Association for Relativistic Dynamics,  June 24 - 26, 2002 at Howard University.
A fuller account of the work is in preparation.}
}

\author{James Baugh$^{a}$, 
David Ritz Finkelstein$^{a}$, 
Andrei Galiautdinov$^{b}$, 
Mohsen Shiri-Garakani$^{a}$\\ \\
$^{a}$ {\small \it School of Physics,
Georgia Institute of Technology, Atlanta, GA
30332-0430}\\
$^{b}$ {\small \it Department of Mathematics and Science,
Brenau University, Gainesville, GA 30501}   }

\date{}

\maketitle

{\abstract {Segal proposed ultraquantum
commutation relations
with two ultraquantum constants
$\hbar', \hbar''$ 
besides Planck's
quantum constant $\hbar$
and with a variable $i$.
The Heisenberg
quantum algebra is a contraction ---
in a more general sense than that of In\"on\"u and Wigner
---
of the Segal ultraquantum algebra.
The usual constant $i$ arises as a vacuum order-parameter
in the quantum limit  $\hbar', \hbar''\to 0$.
One physical consequence is a discrete spectrum
for canonical variables and space-time coordinates.
Another
is an interconversion of time and energy
accompanying space-time meltdown (disorder),
with a fundamental conversion factor
of some kilograms of energy per second.}}


\section{Segal Doctrine}

A group that is  not semisimple we call {\em compound}\/;
a theory based on a compound group  we also call compound.
Compound theories have idols
in the sense of Francis Bacon,
false absolutes,
which couple into other entities
with no reciprocal coupling.

A group is unstable if any neighborhood 
of its structure tensor 
--- defining the product on its Lie algebra --- contains
structure tensors
of a non-isomorphic group \cite{Segal51}.
Then almost every group (structure tensor)
in the neighborhood 
is non-isomorphic to the unstable group.
Since measurements of structure tensors have error bars,
they assign probability 0 to
unstable groups.
Unstable groups in physics are not
based on experimental results as much as on
 faith in an idol.

Lie products $\x:  V\ox V\to V$ 
--- also called structure tensors --- form a quadratic submanifold  $\Xi$
of the space of tensors $V\ox V\to V$.
By a {\em contraction} of a Lie algebra we mean the endpoint
$\x(0)$
of a homotopy $\x(p)$ in $\Xi$
with $\x(1)=\x$, $\x(p)\cong \x$ for $0< p\le 1$, 
and $\x(0)\not\cong \x$.
The concept is Segal's,
the nomenclature ours.
Segal's concept includes
the In\"on\"u-Wigner contraction
as a very special case.

Segal \cite{Segal51} proposed that 
groups of a physical theory should be stable.
He further pointed out that stability requires
semisimplicity:
Compounds are unstable.
The Segal  Doctrine suggests that any
compound physical theory is a contraction
of a more stable, more accurate, semisimple theory,
which we call 
its {\em expansion}.

\section{Examples}

The standard example
of group contraction and expansion is the 
Galileo group, in which 
\BEq
B_x \x B_y = 0, \quad R_z \x B_x = B_y, \quad R_z \x B_y=-B_x,\quad \dots
\EEq
Here $B_x$ is the boost along the $x$ axis,  $R_z$ 
is the rotation about the $z$ axis, and $A\x B:=[A, B]= AB-BA$
is a Lie product.
The Galileo algebra is 
unstable, its idol the absolute time $t$.
Its familiar expansion is the 
Lorentz group, in which now
\BEq
B_x \x B_y = c^{-2} R_z, \quad R_z \x B_x = B_y, \quad R_z \x B_y=-B_x,\quad \dots\/.
\EEq
This expansion is now stable against itself: a further small change in $c$
makes no difference to the group.
The expansion parameter is
$1/c^2 \to 0$.

Every bundle in physics implies a  non-reciprocity,
an idol, an instability, and a compound group.
Indeed, the Galileo group is a bundle, 
the Lorentz is not.

For example, the point $p=(x,y)$ is a simple object,
the chord $(p_1, p_2)$ is a semisimple object,
the Cartesian product of two simples,
but the limiting chord,
the tangent vector $(p, dp)$, is a compound,
containing the simple $p$ but not as a factor
in a Cartesian product.
The differential calculus is thus unstable,
and so may be an unsuitable language
for a supposedly empirical fundamental  physics.

In consequence the Heisenberg Lie algebra defined by
\BEq
p\x x =-\hbar i, \quad \hbar i \x p =x\x  \hbar i=0
\EEq
is not simple or semisimple but compound and unstable. 
Its
idol is $i$.
In the seminal paper \cite{Segal51}
that stimulated the paper of In\"on\"u and Wigner
\cite{Inonu52}, \cite{INONU},
Segal proposed an  expansion
that simplified the Heisenberg algebra.
For homogeneity we introduce
skewsymmetric operators $\wh{x}:= ix$,
 $\wh{p}=-ip$ to go with $i$
and
designate their ultraquantum expansions by
$\breve{x}, \breve{p}, \breve{i}$\/.
The Segal ultraquantum commutation relations are essentially
\BEq
\breve{p}\x \breve{x}=-\hbar \breve{i}, 
\quad \breve{i}\x \breve{p} =-\hbar' \breve{x}, 
\quad \breve{x} \x  \breve{i}=-\hbar''\breve{p}.
\EEq
The compound Heisenberg algebra
has become the simple $\SO(3)$ algebra.
($\SO(2,1)$ is another possibility, 
more promising in several respects.)
This ultraquantum theory has two ultraquantum constants
$\hbar', \hbar''$ in addition to the quantum constant $\hbar$.
It also goes beyond the quantum theory in that it quantizes ---
gives a discrete spectrum to ---
the canonical variables themselves.
The space and time coordinates
obey essentially the same commutation relations
with the momentum and energy,
and therefore
undergo the same expansion
and quantization.

\section{Quantum Principles}

By the relativity group \cite{DRF96} of a system we mean the
group of all admissible frame transformations.
In classical mechanics this is the canonical group of the phase space.
In quantum theory it is 
the unitary group of the Hilbert space of the quantum
system. The Segal stability criterion suggests a fundamental
quantum principle \cite{GF}:

{\em The relativity group of the system is
a simple Lie group.}

We recover the part of quantum logic
expressed by  the ortholattice of  predicates or projection operators
as the lattice of simple Lie subgroups
of the relativity group of the system;
the orthogonality of two predicates is the commutativity
of two orthogonal subgroups, element by element.
It is well-known how to build the rest of the standard
quantum kinematics on this ortholattice foundation.
The ortholattice in question being non-distributive,
the simplicity principle
implies all the well-known quantum paradoxes.

Just as one may associate the unit energy intervals of the
harmonic oscillator with an elementary quantum
or phonon,
we  identify the unit intervals  of the discrete 
ultraquantum spectrum of time
with an elementary operation
which we call the chronon
 \cite{GALIJTP, GF},
characterized by a fundamental unit of time $\Chi$ (Chi).

\section{Standard Model}

We apply the Segal criterion to the Standard Model.
The elements of the theory are
\begin{itemize}
\item Complex numbers {\1C}

\item Charges:  hypercharge, isospin, color $\0U(1)\x \SU(2)\x \SU(3)$

\item Spin $\0S\0L(2, \1C)$

\item {\em Space-time}

\item Fermions

\item  {\em Bosons}

\end{itemize}

We have italicized the unstable compounds.
The instabilities both arise from 
the Heisenberg commutation relations:
for $x$ and $\partial_x$ in the case of space-time,
for annihilators $\phi$ and creators 
$\phi^{\dag}$ in the case of Bose statistics.
The key idol is the $i$ (or 1) on the right-hand side
of the canionical commutation relation.
The fermion algebra is a Clifford algebra,
which is stable.

The diffeomorphism group of general relativity
is likewise unstable
and has the same idol.
Thus present physics
is infested with a cluster of idols all of the same tribe.

The expansion parameters  $\hbar', \hbar''$
 stabilize all these compounds if we 
replace every occurrence of the unstable Heisenberg quantum
commutation relations by the stable
ultraquantum  commutation relations of Segal.
Stability thus leads us to
a real quantum theory rather than complex,
with a large
orthogonal group rather than a unitary one.
The representations of this
group may be found within its Clifford algebra.

Associativity is just as destabilizing
as commutativity.
It seems  likely that experiment 
will eventually
reveal non-associative processes
underlying the present associative ones.
Physically, non-associativity is a kind of binding.
A powerful language for non-associative combination
is set theory.
There binding is represented by the
Cantor brace $\{x\}$, 
which maps any set to a unit set.
This defines
a non-associative product: $\{\{xy\}z\}\ne
\{x\{yz\}\}$. 
We do not pursue non-associative physics at present.
We introduce an 
ultraquantum set theory only
to expand the classical set theory 
incorporated in the present theory of space-time.

We form the mode vector space of ultraquantum set theory
by
expanding  
the mode vector space
of Fermi-Dirac statistics,
a Grassmann algebra.
The result is a free Clifford  algebra $\CLIFF V$ over a 
quadratic space $V$.
We write this as an exponential
\BE
\CLIFF V={\22}^V,
\EE
the quantum correspondent of the  classical power-set exponential
\BE
\0P X=2^X\/.
\EE

We designate the Clifford algebra over a 
quadratic space with dimension $n=n_++n_-$
and signature $s=n_+-n_-$ by $\Cliff(n_+, n_-)$.
Our $\Cliff(n_+, n_-)$ has $n_+$ generators $\ga$
with $\ga^2=-1$ and $n_-$ with $\ga^2=+1$.
In particular
$\Cliff(1,0)=\1C$.

In principle we need not give a separate meaning 
to the sub-symbol $\22$,
since we use it only in the combination $\22^V$.
In fact  we think of $\22$ as a 
two-dimensional Clifford algebra, either the complex numbers
$\Cliff(1,0)$
or the duplex numbers
$\Cliff(0,1)$, depending on the signature of $V$\/.
${\22}^V$ is a Clifford product of $\22$'s,
complex or duplex,
one for each dimension of $V$,
with all its $i$'s anticommuting.
For even dim $V$, we designate the underlying spinor module by
$\Si:=\sqrt{\22}^V$\/.
This power notation for Clifford algebras and spinors
incorporates several useful combinatorial identities. 

\section{Ultraquantum Dirac Algebra}

We unify 
in one Clifford algebra $\Cliff(3,3)^N$ of dimension $2^{6N}$
 the variables of
space-time-momentum-energy-spin
and the
 imaginary
unit $i$ \cite{GALIJTP, GF},
a six-fold unity,
putting off for the moment the unitary charges.
We index Clifford generators
${\ga}^{\om}(n)\in \Cliff(3N,3N)\sim\Cliff(3,3)^N$, 
$\om=1,\dots, 6$, $n=1,\dots,N$.
The index $n$ enumerates processes along
the world line, a proper time variable.
The usual Dirac spin $\ga^{\mu}$
is the ``growing tip''  of the world line, $\ga^{\mu 5}(N)$\/,

The ultraquantum variables
for one spin-1/2 particle are then
\begin{eqnarray}
\breve i  &:=& \frac{1}{N} \sum_{n=1}^{N}
\gamma^{56}(n) ,\cr\cr
\breve x^\mu &:=& -\Chi
\sum_{n=1}^{N}
\gamma^{\mu 5}(n) , \cr\cr
\breve p^\nu
 &:=& {\Ep}
\sum_{n=1}^{N}
\gamma^{\nu 6}(n),\cr\cr
 \tilde \gamma^{\mu} &:= &-\gamma^{\mu 5}
(N).
\end{eqnarray}

In the contraction to the quantum limit of the ultraquantum theory,
 $N\to \infty, $
$\Chi, {\Ep} \to 0,$ 
$i^2 \approx -1,$
we impose $ \Chi {\Ep} N \equiv {\hbar\over 2}$.
Then 
Alg$(\breve i,\breve x, \breve p)$  contracts
to the Heisenberg algebra 
Alg$( i,x, p)$
and
Alg$(\breve i,\breve x, \breve p, \tilde \ga)$ 
contracts to  Dirac's extension of the Heisenberg algebra,
Alg$( i,x, p, \ga)$.

\section{Ultraquantum Dirac Equation}

A natural expansion of the Dirac equation,
with expanded symmetry group, is \cite{GALIJTP} 
\BEqA
\label{DYNAMICS}
\tilde{D}\psi&=&0\cr
\tilde{D}& :=& \frac{2{\Ep}}{\hbar^2} \; \breve
S^{\omega\rho}
   \breve L_{\omega \rho} \cr
& \to& \ga^{\mu}\pa_{\mu}+m_{\Chi}
\mbox{\rm \quad as } N\to \infty, {\Ep}\to 0, {\rm and} \; \Chi
\; {\rm remains} \; {\rm finite }. 
\EEqA
Here,
\BEqA
\breve S^{\omega \rho}&:=& \frac{\hbar}{2}\,
\gamma^{\omega \rho}(N+1),\cr
\breve L_{\omega \rho}&:=& \frac{\hbar}{2}
\sum_{n=1}^{N}
\gamma_{\omega \rho}(n),
\EEqA
with $\omega , \rho = 1, \dots , 6,$ and
$\mu,\nu = 1, \dots, 4$.

We guess that
$m_\Chi\sim $ top-quark mass and
$N\Chi\lo$ age of universe:
\begin{eqnarray*}
m_{\Chi}&\sim &10^2 \GeV \cr
 \Chi&\sim& 10^{-25} \sec \cr
N &\lo &10^{41}.
\end{eqnarray*}

\section{Cosmos Theory}

Quantum theory treats of sharp filtration processes.
No experimenter can carry out sharp filtration
processes on the entire cosmos.
How can one make a sensible cosmos theory?

Actually field theorists do this all the time.
Laplace already did it.
Their method was to ignore the
philosophical problem and invent
an imaginary exocosmic observer.
We do the same.

The Segal criterion helps here:
The purely operationalist theory
is naive, and makes the
absolute experimenter into a destabilizing
idol.

Thus the first step is to renounce operationalism.
Instead of the usual idealized version of the physical
experimenter
$\mathbf{\widehat{S}}$
 on the system 
we introduce an imaginary
Cosmic Experimenter 
$\mathbf{\widehat{U}}$
who  inputs, propagates, and 
outtakes 
the cosmos;
a quantum version of Laplace's fictitious 
omniscient intellect.
The CE does experiments 
described by unitary operators $U$
with matrix elements
of the form
\BEq
\bra \tau=+T|U|\tau=-T\ket,
\EEq
where $2T$ is the duration of the experiment in
the proper time of the CE.
We imagine $\mathbf{\widehat{U}}$ measures
what
$\mathbf{\widehat{S}}$ measures;
and ignores
(traces over) what $\mathbf{\widehat{S}}$ ignores:
especially the proper time $\tau$ of the CE.  
Then cosmic energy
$E=(i/\hbar) d/d\tau$ is a central (``superselection'')
operator.

\section{Q/Q Kinematics}

Field theory began with classical fields on classical space-time (C/C),
evolved to quantum fields on classical space-time (Q/C),
and we can now formulate a Q/Q theory (quantum fields on quantum space-time).

Question: How do we define a  field space
\BEq
F=f^b
\EEq
with quantum fiber and base defined by mode spaces  $f, b$? By correspondence we
demand that these spaces admit a concept of dimension with
$\Dim f^b = (\Dim f)^{\Dim b}$.

Answer \cite{fs}: Set $f=(\surd\22)^{\phi}$.

Then 
\BE
\label{eq:F}
F=f^b := ((\surd\22)^{\phi})^b =(\surd\22)^{\phi\ox b}.
\EE
For this construction the field $f$ must be spinorial, of course.  So it is.

\section{Qubits to Qunats}

 We propose  to use the new algebraic
structure created by the $\Chi$ expansion
to make the observed bosons out of our fundamental
fermions. 
Fermion occupation numbers are quantum binary numbers
with eigenvalues 
$b=0, 1$: {\em qubits}.
A boson occupation number is a variable natural number 
with eigenvalues
$n=0, 1, 2, 3, \dots$:  a {\em qunat}.
We have to make qunats out of qubits.

It is easy to make ``nats'' (natural numbers)
out of bits;
evidently 1 nat = $\infty$ bits.
Likewise 1 qunat = $\infty$ qubits.
Here is one way of making a qunat $(q,p)$ from $N$ qubits
$\ga(i,n)$, 
$i=0,1,2, \quad 0\le n \le N$
in the limit $N\to \infty$:
\begin{eqnarray}
\widehat{q} &=& Q\;\Si_n \ga^{01}(n),\cr
\widehat{p} &=&P \;\Si_n \ga^{02}(n),\cr
\widehat{\imath} &=&N^{-1}\;\Si_n \ga^{12}(n),\cr
\widehat{q} \x \widehat{p}&=&\widehat{i},\cr
\widehat{\imath} \x \widehat{q}&=&\hbar'\widehat{p},\cr
\widehat{p} \x \widehat{\imath}&=&\hbar''\widehat{q},\cr
\widehat{\imath}^2&=&-1.\cr
2 NQP&=&1 , \quad 2 Q=N \hbar, \quad 2P Q=\hbar'.
\end{eqnarray}
Here $\hbar',\hbar'' << 1$
are Segal ultraquantum constants and
the value of the qunat is
the usual oscillator Hamiltonian
$H={\om\over 2} (p^2+q^2) - H_0$
with minimum eigenvalue $H_0$ reset to 0
by subtraction.
For finite $J, Q, P$, however --- and this is
supposedly the actuality ---
these are not oscillators but rotators,
namely spins,
and the nat is bounded by $2l+1$.

\section{Covariant Finite Toy Q/Q Kinematics}

The above $\Cliff(3,3)$ is a toy:
the least Clifford algebra expanding and stabilizing
the Poincar\'e and Heisenberg Lie algebras.
It doesn't accommodate the standard model.
In quantum set theory,
after the Dirac algebra $C^{2}=\Cliff(1,3)$
comes $C^{3}=\Cliff(10,6)$,
a Clifford algebra 
over a 16-dimensional space instead of a 6-dimensional one.
The additional 10 dimensions might accommodate the GUT,
and 16 resonates with the Cartan-Atiyah-Singer-Botts
octal periods. 

Let us apply the Q/Q field construction (\ref{eq:F}) to the $\Cliff(3,3)$  toy:
 \begin{eqnarray*}
\phi&=&\surd \Cliff(3,3)\sim\28 ,\cr
f&=&(\surd\22)^{\phi}\sim (\surd{\22})^{8}\sim \21\26.
\end{eqnarray*}
The number of qubits per cell in this toy is 16.
The main thing is that the number is finite and the theory is covariant.

\section*{ACKNOWLEDGMENTS}

One of the authors (A.G) acknowledges support from grant no.
NICHD HD39787-02.
We thank Larry Horwitz for helpful suggestions.


\begin{thebibliography}{}



\bibitem{fs}
J. Baugh, D. Finkelstein, A. Galiautdinov, and H.
Saller, {\it J. Math. Phys.} {\bf 42}, 1489 (2001)


\bibitem{DRF96}
D. Finkelstein, {\it Quantum Relativity}
(Springer-Verlag, New York, 1996)


\bibitem{FG}
D. Finkelstein and A. Galiautdinov, {\it J.
Math. Phys.} {\bf 42}, 3299 (2001).

\bibitem{GALIJTP}
A.A. Galiautdinov, {\it IJTP} {\bf 41}, 1423 (2002)

\bibitem{GF}
A.A. Galiautdinov and D.R. Finkelstein, {\it J.
Math. Phys.} {\bf 43}, 4741 (2002)


\bibitem{INONU}
E. In\"on\"u, Contraction of Lie Groups and their
Representations, in F. G\"ursey, ed., {\it Group Theoretical
Concepts and Methods in Elementary Particle Physics}, pp.
391--402 (Gordon and Breach, Science publishers, New York,
1964)

\bibitem{Inonu52}
E. In\"on\"u and E.P. Wigner, {\it Proc.
Nat. Acad. Sci.} {\bf 39}, 510 (1952)

\bibitem{NW}
C. Nayak and F. Wilczek, {\it Nucl. Phys.}
{\bf B479}, 529 (1996)


\bibitem{Segal51}
I.E. Segal,
{\it Duke Math. J.} {\bf 18}, 221 (1951)


\bibitem{WILCZEK}
F. Wilczek,
{\it Nucl.Phys.Proc.Suppl.} {\bf 68}, 367 (1998).
Also hep-th/9710135.

\bibitem{wilczek}
F. Wilczek, hep-th/9806228 [LANL]

\end{thebibliography}
\end{document}